\documentclass[aps,prl,twocolumn,superscriptaddress, notitlepage]{revtex4-1}
\usepackage{hyperref}
\usepackage{graphicx}
\usepackage{amsmath}
\hypersetup{colorlinks=true,linkcolor=blue,citecolor=blue}

\begin{document}

\title{ Contact-based and spheroidal vibrational modes of a hexagonal monolayer of microspheres on a substrate}
\author{A.Vega-Flick }
\affiliation{Department of Chemistry, Massachusetts Institute of Technology, Cambridge, Massachusetts 02139, USA}
\affiliation{Applied Physics Department, CINVESTAV-Unidad M\'erida, Carretera Antigua a Progreso Km 6, Cordemex, M\'erida, Yucat\'an,  97310, M\'exico}
\author{R. A. Duncan}
\affiliation{Department of Chemistry, Massachusetts Institute of Technology, Cambridge, Massachusetts 02139, USA}
\author {S. P. Wallen} 
\author{ N. Boechler}
\affiliation{Department of Mechanical Engineering, University of Washington, Seattle, WA, 98195 USA}
\author{C. Stelling}
\author{M. Retsch}
\affiliation{Physical Chemistry, University of Bayreuth, Universitaetsstr 30, 95447 Bayreuth, Germany}
\author{J.J. Alvarado-Gil}
\affiliation{Applied Physics Department, CINVESTAV-Unidad M\'erida, Carretera Antigua a Progreso Km 6, Cordemex, M\'erida, Yucat\'an, 97310, M\'exico}
\author{K. A. Nelson}
\affiliation{Department of Chemistry, Massachusetts Institute of Technology, Cambridge, Massachusetts 02139, USA}
\author{A. A. Maznev}
 \affiliation{Department of Chemistry, Massachusetts Institute of Technology, Cambridge, Massachusetts 02139, USA}

\begin{abstract}
We study acoustic modes of a close-packed hexagonal lattice of spheres adhered to a substrate, propagating along a high-symmetry direction. 
The model, accounting for both normal and shear coupling between the spheres and between the spheres and the substrate, yields three 
contact-based vibrational modes involving both translational and rotational motion of the spheres. Furthermore, we study the effect of 
sphere-substrate and sphere-sphere contacts on spheroidal vibrational modes of the spheres within a perturbative approach. The sphere-substrate 
interaction results in a frequency upshift for the modes having a non-zero displacement at the contact point with the substrate. Sphere-sphere 
interactions result in dispersion of spheroidal modes turning them into propagating waves, albeit with a small group velocity. Analytical 
dispersion relations for both contact-based and spheroidal modes are presented and compared with results obtained for a square lattice.

\end{abstract}
\maketitle

\section{I. Introduction}

Acoustic wave propagation in ordered granular materials has become an increasingly active area of research. 
This is due in part to a wide array of exhibited acoustic phenomena stemming from the Hertzian interaction between 
the particles and the periodic particle arrangement \cite{Nesterenko2001, Theocharis2013}. Modern colloidal assembly techniques provide 
the ability to fabricate 
2D granular structures of micro- to nanometer-sized spherical particles \cite{Vogel2015}. Recently, laser-based techniques were used to study 
acoustic waves in self-assembled 2D microgranular monolayers \cite{Boechler2013, Khanolkar2015, Hiraiwa2016, Eliason2016}. It was shown that 
adhesion between the particles and between the particles and the substrate plays an important role in 
determining the acoustic properties of these particulate assemblies \cite{Boechler2013, Khanolkar2015, Hiraiwa2016, Eliason2016}. 
A theoretical analysis of acoustic waves in a 2D granular crystal was initially limited to free-standing granular membranes \cite{Tournat2011}. 
A further study \cite{Wallen2015} accounted for the effect of the substrate present in the experiments. The existence of three contact-based vibrational 
modes involving both translational and rotational motion of the spheres predicted in Ref. \cite{Wallen2015} was subsequently confirmed in the 
experiment \cite{Hiraiwa2016}. However, the analysis \cite{Wallen2015} was performed for a square lattice of spheres, whereas  experimental 
studies used close-packed hexagonal microsphere monolayers \cite{Boechler2013, Hiraiwa2016}. 

Thus one objective of the present study is to extend the analysis of contact-based vibrational modes  of a granular monolayer on a substrate 
onto the case of a hexagonal lattice. We will see that the behavior of contact-based modes propagating in a high symmetry direction of a 
hexagonal lattice is qualitatively similar but quantitatively different compared to the square lattice case. Our further objective is to analyze intrinsic 
spheroidal vibrations in a microgranular monolayer. Such 
spheroidal vibrations have been experimentally observed in both 3D \cite{Cheng2006} and 2D \cite{Khanolkar2015} microgranular  assemblies. We 
employ a perturbation approach to study the effects of particle-particle and 
particle-substrate contacts on the spheroidal modes. Our analysis involves solving two problems: (i) the effect of the substrate on spheroidal 
modes of an individual sphere; (ii) the effect of sphere-sphere contacts which transform spheroidal modes of individual spheres into collective 
propagating modes. The former problem was considered previously \cite{Tian2004} in the context of resonant ultrasound 
spectroscopy of macroscopic spheres. Our analysis is similar to the quasi-static contact model presented in Ref. \cite{Tian2004}; however, 
we employ an efficient energy-based perturbation approach leading to a simple explicit expression for the substrate-induced frequency shift. 
Furthermore, our analysis includes spheroidal modes with horizontal displacement at the contact point with the substrate, which were not considered 
in Ref. \cite{Tian2004}. The second problem was previously analyzed numerically, using a finite element method, for a linear chain 
of joined spheres \cite{Hladky-Hennion2002, Hladky-Hennion2004}; it was found that the dispersion of the collective 
vibrational modes is similar to that of a prototypical chain of weakly-coupled oscillators \cite{Hladky-Hennion2004}. We present 
an analytical energy-based perturbation analysis yielding simple explicit dispersion equations. Our analysis of contact-based and spheroidal 
modes will be illustrated by results obtained for a monolayer of micron-sized silica spheres on a silica substrate. 

\section{II. Contact-based modes}

\begin{figure}[h]
\centering
\graphicspath{ {./} }
\includegraphics[width=240pt]{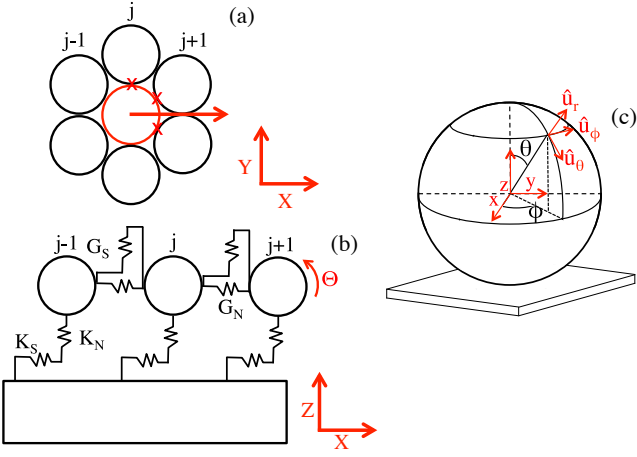}
\caption{(a) Top down view of the hexagonal arrangement of microspheres. Red crosses indicate the contact springs of the unit cell 
considered in the calculation of the sphere-sphere contact dispersion, arrow indicates the wave propagation direction. 
(b) Schematic representation of the dynamic model of contact-based modes. (c) Schematic diagram regarding the spheroidal vibrational modes.}
\label{fig1}
\end{figure}
We consider a close-packed hexagonal monolayer of spheres of diameter $D$ assembled on a substrate. We follow the approach of Ref. \cite{Wallen2015}, 
modifying it for the case of a hexagonal lattice. Both the spheres and the substrate are considered rigid, i.e., internal elastic vibrations of the spheres as 
well as elastic waves in the substrate are 
disregarded. However, we account for the local elasticity of the spheres and the substrate at the contact point by employing the Hertz-Mindlin model of 
an elastic contact \cite{Johnson1987}. The sphere-substrate 
and sphere-sphere contacts are modeled as normal and shear springs. The spring constants $K_{N}$, $K_{S}$ correspond to sphere-substrate contact 
and $G_{N}$, $G_{S}$ to sphere-sphere contact, where the subscript $N$ relates to normal contact stiffness and $S$ to shear contact stiffness (see 
Fig. \ref{fig1}(b)). The spring constants are calculated using Hertz-Mindlin and the DMT (Derjaguin-Muller-Toporov) model \cite{Johnson1987, Muller1983}. In this case, 
the normal and shear spring constants are not independent of each other; their ratios are determined by the elastic constants of the contacting 
materials, such that 

\begin{equation}
\label{eq1}
\begin{aligned}
& \frac{K_{N}}{K_{S}}= \frac{E_{2}(\nu_{1}-2)(\nu_{1}+1)+E_{1}(\nu_{2}-2)(\nu_{2}+1)}{2 E_{2}(\nu_{1}^{2}-1)+2 E_{1}(\nu_{2}^{2}-1)}, \\
& \frac{G_{N}}{G_{S}}=\frac{(2-\nu_{1})}{2(1-\nu_{1})}, 
\end{aligned}
\end{equation}

\noindent where $\nu$ is the Poisson ratio and $E$ is the Young's modulus. Subscripts 1 
and 2 correspond to sphere and substrate, respectively. 

\noindent Each sphere has six degrees of freedom (three translational and three rotational), giving 
rise to six contact-based vibrational modes of the monolayer. We assume that the wavevector is in the high-symmetry direction shown in Fig. \ref{fig1}(a), 
which corresponds to the $\Gamma$-K direction of the reciprocal lattice. In this case, three out of six vibrational modes involve the motion of the spheres 
confined to the sagittal plane, containing the wavevector and the surface normal. In the present work, we are only interested in these sagittally polarized 
modes, as they can be excited and probed in laser-based experiments \cite{Boechler2013, Khanolkar2015, Hiraiwa2016}.

Following the procedure by Wallen et al. \cite{Wallen2015}, we write the equations of motion for the j$^{th}$ sphere considering waves propagating 
in the $\Gamma$-K direction 
of the microsphere lattice,
as shown in Fig. \ref{fig1}(a):

\begin{equation}
\footnotesize
\label{eq2}
\begin{aligned}
& \textrm{m}\ddot{Z_{j}}=-K_{N}Z_{j}+ \\
 & 2G_{S}\bigg[ (Z_{j+1} -2Z_{j}+Z_{j-1})-\frac{\sqrt{3}}{2}R(\Theta_{j+1}-\Theta_{j-1})  \bigg] ,\\
& \textrm{m}\ddot{X_{j}}=-K_{S}(X_{j}+R\Theta_{j})+ \\
& (\sqrt{3}G_{N}+G_{S})\bigg[ (X_{j+1} -2X_{j}+X_{j-1})  \bigg], \\
& I\ddot{\Theta_{j}}=-K_{S}R(X_{j}+R\Theta_{j})+ \\
& \sqrt{3}G_{S}R\bigg[(Z_{j+1}-Z_{j-1}) - \frac{\sqrt{3}}{2}R(\Theta_{j+1}+2\Theta_{j}+\Theta_{j-1})  \bigg], 
\end{aligned}
\end{equation}

\noindent where $R=D/2$ is the sphere radius,  m is the sphere mass, $I$ is the moment of inertia (for a solid sphere $I=(2/5)\textrm{m}R^{2}$) 
and the vertical, horizontal, and angular 
displacements of the j$^{th}$ sphere are given by Z$_{j}$, X$_{j}$, and $\Theta_{j}$, respectively. 

Assuming a spatially discrete solution of the form $\hat{Z}e^{iq(2R)j-i\omega t}$ (with similar terms for the other sphere displacements), we obtain 
a system of three linear equations, leading to the following dispersion relation of the monolayer 

\begin{equation}
\label{eq3}
\arraycolsep=8pt
 \begin{vmatrix}
a_{11}       & 0 & a_{13}  \\
    0      & a_{22} & a_{23} \\
a_{31}       & a_{32} & a_{33}
\end{vmatrix} =0 
\end{equation}

\begin{equation}
\footnotesize
\label{eq4}
\begin{aligned}
& a_{11}=\textrm{m}\omega^{2}-K_{N}-4G_{S}\big(1-\cos(qR\sqrt3)\big), \\
& a_{13}=-i2\sqrt{3}G_{S}\sin(qR\sqrt{3}), \\
& a_{31}=-a_{13}, \\
& a_{22}=\textrm{m}\omega^{2}-K_{S}-2\big(\sqrt{3}G_{N}+G_{S}\big)\big(1-\cos(qR\sqrt3)\big), \\
& a_{23}=a_{32}=-K_{S}, \\
& a_{33}=\frac{I\omega^{2}}{R^{2}}-K_{S}-3G_{S}\big(1+\cos(qR\sqrt{3})\big).
\end{aligned}
\end{equation}

\begin{figure}[h]
\centering
\graphicspath{ {./} }
\includegraphics[width=200pt]{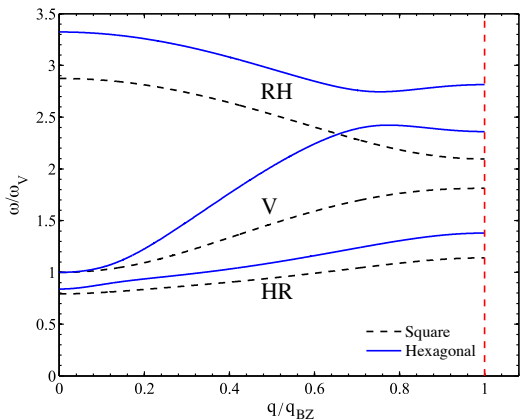}
\caption{Dispersion relations of contact-based modes for hexagonal and square lattices. The frequency axis is normalized to 
the frequency ($\omega_{V}=(2\pi)144$ MHz) of the V mode in the limit $q \rightarrow 0$. The wavevector axis is normalized to the BZ boundary ($q_{BZ}$) of 
the respective hexagonal or square lattice.}
\label{fig2}
\end{figure}

\subsection{A. Dispersion relation analysis}

For the numerical examples discussed in this paper, we choose silica spheres of $D=1$ $\mu$m in diameter on a 
silica substrate. The work of adhesion is $0.063$ J/m$^{2}$ \cite{Israelachvili2011}. The silica properties used in this work 
are $E=73$ GPa (Young's modulus), $\nu=0.17$ (Poisson's ratio)  and $\rho=2.2$ g/cm$^{3}$ (density). 
This results in contact stiffnesses with values of $K_{N}=943$ N/m, $K_{S}=855$ N/m, 
$G_{N}=594$ N/m and $G_{S}=538$ N/m.

\begin{figure}
\centering
\graphicspath{ {./} }
\includegraphics[width=240pt]{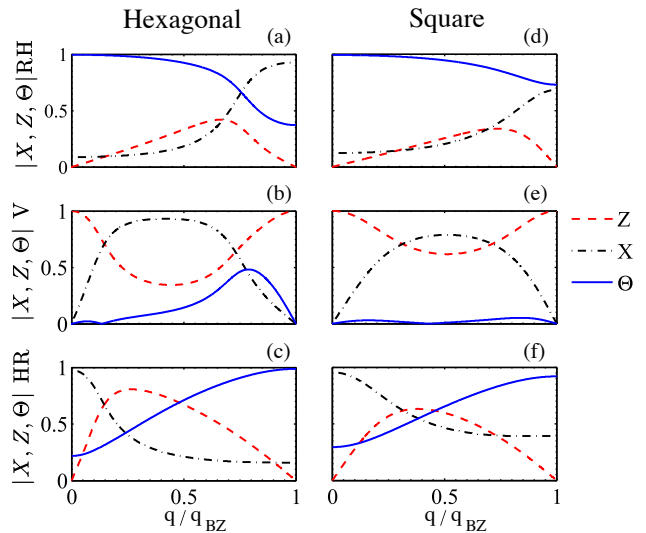}
\caption{Relative amplitudes of the displacement variables for each branch (RH, V, and HR) of the contact-based vibrational modes. 
(a)-(c) Hexagonal lattice. (d)-(f) square lattice.}
\label{fig3}
\end{figure}

Figure \ref{fig2} shows the calculated dispersion relations for both the hexagonal and square microsphere lattices whereas Fig. \ref{fig3} shows the relative amplitudes of displacements and rotations in the acoustic modes. Qualitatively, hexagonal and square lattices yield a similar behavior. In the long wavelength 
limit ($q=0$), one mode corresponds to exclusively vertical motion of the spheres (referred to as ``V'') with a resonant frequency of $\omega_{V}=\sqrt{K_{N}/\textrm{m}}$, 
while the other two modes consist of a combination of 
horizontal and rotational motion; one predominantly horizontal (referred as ``HR'') and the other predominantly rotational (referred as ``RH''). The 
same is found to be the case for a hexagonal lattice. As shown in Fig. \ref{fig3}, these motion patterns change across the BZ, with more changes 
observed in for the hexagonal lattice. For example, the V mode becomes predominantly horizontal at $q/q_{BZ}\sim0.5$, and aquires a significant 
rotational component at higher wavevectors.  

\subsection{B. Limiting cases}
\begin{figure}
\centering
\graphicspath{ {./} }
\includegraphics[width=200pt]{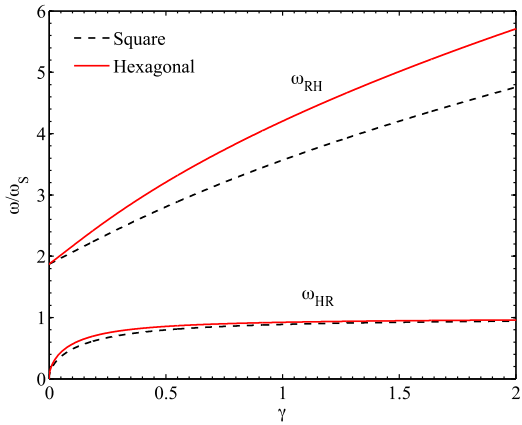}
\caption{Frequencies $\omega_{HR}$ and $\omega_{RH}$ as functions of $\gamma$ in the limit $q=0$, for hexagonal and square lattices. Frequencies 
were normalized to the horizontal resonance frequency $\omega_{S}=\sqrt{K_{S}/m}$. }
\label{fig4}
\end{figure}

The limiting case of long wavelengths ($q\rightarrow0$), the V mode tends to the resonant frequency of $\omega_{V}$, just as in the case of a 
squared lattice \cite{Wallen2015}, as the microsphere monolayer is 
undergoing only vertical motion independent of the lattice configuration. The HR and RH modes have the following frequencies at $q=0$,

\begin{equation}
\label{eq5}
\begin{aligned}
& \omega^{hex}_{HR}=\sqrt{\bigg(\frac{K_{s}}{4m}\bigg)   \big(30\gamma+7-\sqrt{900\gamma^{2}+180\gamma+49}\big)},\\ 
& \omega^{hex}_{RH}=\sqrt{\bigg(\frac{K_{s}}{4m}\bigg)   \big(30\gamma+7+\sqrt{900\gamma^{2}+180\gamma+49}\big)},
\end{aligned}
\end{equation}

\noindent where $\gamma=G_{S}/K_{S}$. These values are different from those for a square lattice found in Ref. \cite{Wallen2015}. We find that the 
HR and RH frequencies for the hexagonal lattice are always higher compared to the square lattice, up to a maximum factor of $\sqrt{3/2}$. This is not surprising 
as the hexagonal lattice can be thought of as being ``stiffer'' than the square lattice due to a larger number of nearest neighbors. In the case where no particle 
rotation is present ($I\rightarrow \infty$), $\omega_{HR}=0$ and $\omega_{RH}$ will reduce to the horizontal frequency 
$\omega_{S}=\sqrt{K_{S}/\textrm{m}}$ for both hexagonal and square lattices. Fig. \ref{fig4} shows $\omega_{HR}$ and $\omega_{RH}$ as a function of 
$\gamma$ normalized by $\omega_{S}$.

For wavevectors at the BZ boundary, the expressions for the frequencies of the three modes for the hexagonal and square lattice are 
shown in Appendix A. 

\section{III. Spheroidal vibrations}

Spheroidal vibrational modes of a free elastically isotropic sphere originally considered by Lamb \cite{Lamb1881} can be obtained by solving the 
elastodynamic equations in spherical coordinates (we used the spherical coordinate system 
shown in Fig. \ref{fig1}(c)) with stress-free boundary conditions at the surface \cite{Eringer1975}. Equations defining the displacement patterns and 
frequencies of the spheroidal modes are presented in Appendix B. Each spheroidal mode is characterized by three numbers $L$, $m$, and 
$n$ referred to as polar, azimuthal, and radial respectively. The polar number $L$ is a non negative integer, the azimuthal number has $L+1$ 
integer values ranging from $m=0,...,L$. As in the case of the contact-based modes, we only consider spheroidal modes with the displacement pattern symmetric with respect to the sagittal ($x$-$z$) plane relevant to laser-based experiments, considering 
the constrains imposed by excitation symmetry and detection limitations \cite{Boechler2013, Khanolkar2015, Dehoux2009}. Additionally we will only consider modes with 
radial number $n=0$ (lowest frequency harmonic of a $L$, $m$ mode).

The spheres in the monolayer are in contact with both the substrate and each other. We consider the contact of the substrate and 
neighboring spheres as a small perturbation to the spheroidal 
mode of an isolated free sphere. We use an energy-based perturbation method built on the requirement that time-averaged kinetic 
and potential energies be equal (additional potential energy due to contact spring necessitates a change in the kinetic energy leading 
to a frequency shift).

\subsection{A. Substrate perturbation}
We begin by considering the contact with the substrate. For a free sphere mode $L$, $m$ with vibration frequency $\omega_{0}$, the average 
kinetic and potential energy per oscillation period are

 \begin{equation}
\label{eq6}
\begin{aligned}
<E_{kin}>&=\frac{1}{4}\omega_{0}^{2}M_{L,m} A^{2},\\
<E_{pot}>&=\frac{1}{4}K_{L,m} A^{2},
\end{aligned}
\end{equation}

 \begin{equation}
\label{eq7}
\begin{aligned}
 & M_{L,m} \equiv \frac{\rho}{A^{2}} \int \lvert u\rvert _{r,L,m}^{2} + \lvert u\rvert_{\theta,L,m}^{2}+\lvert u\rvert_{\phi,L,m}^{2}dV, \\
\end{aligned}
\end{equation}

\noindent where $u_{r,L,m}$, $u_{\theta,L,m}$ and $u_{\phi,L,m}$ are the radial, polar and azimuthal spheroidal displacements, 
respectively, and $A$ is the sphere vibration amplitude.
Note that $K_{L,m}$ and $M_{L,m}$ are not the spring constant and the mass of the sphere; 
they are coefficients obtained by calculating the kinetic and potential energy of a given eigenmode. They are related by the 
expression $K_{L,m}=\omega_{0}^{2}M_{L,m}$, and are independent of the amplitude $A$ \cite{Mead1973}. 

If the sphere is in contact with the substrate, we should add the potential energy of the contact springs, 

\begin{equation}
\label{eq8}
\begin{aligned}
 <E_{kin}>&= \frac{1}{4}\omega_{1}^{2}M_{L,m}A^{2}, \\
<E_{pot}> &=\frac{1}{4}K_{L,m} A^{2}+ \frac{1}{4}K_{N}\lvert u\rvert_{z}^{2} +\frac{1}{4} K_{S}\lvert u\rvert_{x}^{2},
\end{aligned}
\end{equation}

\noindent where $u_{z}=-u_{r,L,m}(R,\pi)$, and $u_{x}=u_{\theta,L,m}(R,\pi,0)$ are vertical and horizontal displacements of the sphere at the 
contact point with the substrate, and $\omega_{1}$ is the perturbed frequency. The deformation 
of the contact springs is given by the surface displacement of the 
sphere at the contact point, i.e. at bottom of the sphere (r = R, $\theta=\pi$). Equating the kinetic and potential energies, we get

\begin{equation}
\label{eq9}
\begin{aligned}
\omega_{1}^{2}M_{L,m}=K_{L,m}&+K_{N}\frac{\lvert u\rvert_{z}^{2}}{A^{2}} +K_{S}\frac{\lvert u\rvert_{x}^{2}}{A^{2}}.
\end{aligned}
\end{equation}
\noindent Further simplification leads to the equation for the 
perturbed frequencies of a spheroidal mode $L$, $m$ due to contact with the substrate

\begin{equation}
\label{eq10}
\begin{aligned}
\omega_{1}^{2} =\begin{cases}
\omega_{0}^{2}+\frac{K_{N}}{M_{0}}C_{N}, \quad &\textrm{for} \quad m=0\\
\omega_{0}^{2}+\frac{K_{S}}{M_{0}}C_{S},  \quad &\textrm{for} \quad m=1 \\
\omega_{0}^{2},  \quad &\textrm{for} \quad m >1
\end{cases}
\end{aligned}
\end{equation}

\begin{equation}
\label{eq11}
\begin{aligned}
& C_{N} \equiv M_{0}\frac{\lvert u\rvert_{r,L,0}^{2}(R,\pi)}{ M_{L,0} A^{2}}, \\
& C_{S} \equiv M_{0} \frac{\lvert u\rvert_{\theta,L,1}^{2}(R,\pi,0)}{M_{L,1}A^{2}},
\end{aligned}
\end{equation}
\noindent where $M_{0}$ is the microsphere mass and $C_{N}$, $C_{S}$ are dimensionless factors. 

The three different cases of Eq. \ref{eq10} are a direct consequence of the functional form of the 
associated Legendre polynomials ($P_{L}^{m}(\cos \theta$)) evaluated at the substrate contact. The radial displacement has nonzero 
values if the mode is $m=0$, this is due to the property 
$P_{L}^{m}(\pm1)=0$ for $m\neq0$. For polar displacements, we use the recurrence relation

\begin{equation}
\label{eq12}
\begin{aligned}
 \frac{\partial P_{L}^{m}(\cos{\theta})}{\partial \theta} & = \frac{L \cos{\theta} P_{L}^{m}(\cos{\theta}) -(L+m)P_{L-1}^{m}(\cos{\theta}) }{\sin{\theta}},\\
\end{aligned}
\end{equation}

\noindent where $lim_{\theta\to\pi} \frac{\partial P_{L}^{m}}{\partial \theta}=0$ for $  m  \neq1$ \cite{Arfken2005}, indicating that the polar 
displacement is nonzero for modes $m  =1$. In the case of the azimuthal direction, the displacements are 
proportional to $P_{L}^{m}(\cos{\theta})/\sin{\theta}$. Using Rodrigue's formula \cite{Arfken2005} for the associated Legendre 
polynomials (Appendix C), $u_{\phi,L,m} \propto m \sin(\theta)^{m-1}$, indicating that modes with $ m \neq1$ have zero azimuthal 
displacement at $\theta=\pi$.

The frequency shift experienced by the microspheres depends on the ratio of the potential energy of the contact to the total potential 
energy of the mode. This means that the larger the displacement of the sphere at the contact point (compared to the deformation across the 
entire volume of the sphere), the greater the frequency shift will be. This can be observed in Table \ref{table1}, where we calculated the frequency shifts 
for spheroidal modes with $m=0,1$ of silica spheres on a silica substrate, using Eq. \ref{eq10}. The $L=2$, $m=0$ mode 
shows the largest frequency shift. This is due to the fact that its surface displacement is the largest at $\theta=0$ and $\theta=\pi$. In 
contrast, the $L=1$, $m=0$ mode has almost no surface displacement and is primarily composed of vibrations of the inner part of the sphere. Consequently, 
it has the smallest frequency shift of the $m=0$ modes. 

\begin{figure}[h]
\centering
\graphicspath{ {./} }
\includegraphics[width=240pt]{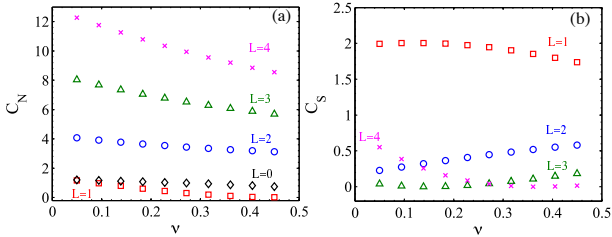}
\caption{(a) Coefficients $C_{N}$ and $C_{S}$ calculated using Eq. \ref{eq11} as a function of $\nu$ for modes with different $L$.}
\label{fig5}
\end{figure}

Modes $ m =1$ undergo smaller frequency shifts, compared to $m=0$. There is an exception in the case of the $L=1$ mode, since the 
amount of polar and azimuthal surface displacements of the $m=1$ mode are larger than the 
almost non existing radial displacement of the $m=0$ mode. Dimensionless contacts $C_{N}$ and $C_{S}$ depend on the polar number $L$ and Poisson's ratio. Figure \ref{fig5} shows $C_{N}$ and $C_{S}$ 
as functions of $\nu$. The size of the sphere also plays an important role in determining the frequency shift. In Eq. \ref{eq10}, $\omega_{0}$ has a $D^{-1}$ dependence, 
the sphere mass goes as $D^{-3}$, and the contact stiffnesses $K_{N}$ and $K_{S}$ have 
a $D^{2/3}$ dependence \cite{Wallen2015}. This results in a $D^{-1/3}$ dependence of the relative frequency shift. Thus the effect is larger for smaller spheres.

\begin{table}[h]
\centering
\caption{\label{table1} Spheroidal frequency shifts due to contact with the substrate for micron-sized silica spheres on a silica substrate. The 
shift in frequency is presented as percent deviation from the free sphere frequencies, whose values are also shown for reference.}
\small
\begin{ruledtabular}
\begin{tabular}{ccccccc}
$L$ & 0& 1& 2 & 3& 4 & 5 \\
$\frac{\Delta \omega}{\omega_{0}} \quad \textrm{for} \quad  m=0 (\%) $ & 0.054 & 0.045 &0.384 & 0.343 & 0.322& 0.308                         \\ 
$\frac{\Delta \omega}{\omega_{0}} \quad \textrm{for} \quad  m=1 (\%)$  & -& 0.124 & 0.033 & 0.00006& 0.005 & 0.014  \\
\hline
$\omega_{0}/2\pi (\textrm{GHz})$ & 4.56 & 3.89 &3.15 & 4.65 &5.92 & 7.12                          
\end{tabular}
\end{ruledtabular}
\label{table1}
\end{table}

\subsection{B. Sphere-sphere perturbation}

We now consider the perturbation due to the sphere-sphere contacts. For the case of the hexagonal lattice, a unit cell 
of the microsphere monolayer contains three contact points shown by the red ``X'' markers in Fig. \ref{fig1}(a). The sphere in the j$^{th}$ 
position is coupled to its neighbours and to the substrate through the normal contact springs $G_{N}$ and $K_{N}$ respectively. We assumed 
that the displacement of the spheroidal modes with wavevector along $x$ is given by 

\begin{equation}
\label{eq13}
\begin{aligned}
u_{L,m} \propto e^{iqx_{j}-i\omega t},
\end{aligned}
\end{equation}

\noindent We set the coordinates of the j$^{th}$ column of spheres to zero, such that $x_{j}$ = 0 and $x_{j+1}=(\sqrt{3} /2)D$. The average kinetic and potential energy of 
the unit cell over an oscillation period can be expressed as

\begin{equation}
\label{eq14}
\footnotesize
\begin{aligned}
 <E_{kin}>&=\frac{1}{4}\omega^{2}M_{L,m}A^{2}, & \\
 <E_{pot}>&=\frac{1}{4}K_{L,m}A^{2}  + \\
	 & \frac{G_{N}}{4}\big(u_{j,r}(R,\frac{\pi}{2},\frac{\pi}{2})+u_{j,r}(R,\frac{\pi}{2},\frac{3\pi}{2}) \big)^{2}  \\
	  & \frac{G_{N}}{2}\big(u_{j,r}(R,\frac{\pi}{2},\frac{\pi}{6})+u_{j+1,r}(R,\frac{\pi}{2},\frac{7\pi}{6})\big)^{2}+\\
	 & \frac{G_{S}}{2}\big(u_{j,\theta}(R,\frac{\pi}{2},\frac{\pi}{6})-u_{j+1,\theta}(R,\frac{\pi}{2},\frac{7\pi}{6})\big)^{2}+\\
	 & \frac{G_{S}}{2}\big(u_{j\phi}(R,\frac{\pi}{2},\frac{\pi}{6})+u_{j+1,\phi}(R,\frac{\pi}{2},\frac{7\pi}{6})\big)^{2}, 
\end{aligned}
\end{equation}

\noindent where $K_{L,m}=\omega_{1}M_{L,m}$. $M_{L,m}$ and $\omega_{1}$ are given by Eq. \ref{eq10} and \ref{eq7} respectively. The 
subscripts $L$, $m$ of the displacements are implied even though the are omitted.
We note the positive sign between the j$^{th}$ and j$^{th+1}$ radial and azimuthal sphere displacements even though the elongation of the 
contact spring should be equal to the difference. This is because the unit vectors $\hat{r}$ and $\hat{\phi}$ evaluated at diametrically opposed positions have 
opposite signs. This is not the case for the polar displacements, where the unit vectors $\hat{\theta}$ at diametrically 
opposed positions have the same sign. In the case of contact between spheres in the same $j^{th}$ column, only the radial displacements 
elongate the contact springs. After further simplification

\begin{equation}
\label{eq15}
\footnotesize
\begin{aligned}
&<E_{kin}> =\frac{1}{4}\omega^{2}M_{L,m}A^{2} \\
&<E_{pot}> =\frac{1}{4}K_{L,m}A^{2}+ \\
&\begin{cases}
G_{N}\big(u_{j,r}^{2}(R,\frac{\pi}{2},\frac{\pi}{2})+u_{j,r}^{2}(R,\frac{\pi}{2},\frac{\pi}{6})[1+\cos qD\sqrt{3}/2]\big)+ & \quad m=\text{even}\\
G_{S}u_{j,\theta}^{2}(R,\frac{\pi}{2},\frac{\pi}{6})[1-\cos qD\sqrt{3}/2]+\\
G_{S}u_{j,\phi}^{2}(R,\frac{\pi}{2},\frac{\pi}{6})[1+\cos qD\sqrt{3}/2],\\
G_{N}u_{j,r}^{2}(R,\frac{\pi}{2},\frac{\pi}{6})[1-\cos qD\sqrt{3}/2]+& \quad m=\text{odd}\\\
G_{S}u_{j,\theta}^{2}(R,\frac{\pi}{2},\frac{\pi}{6})[1+\cos qD\sqrt{3}/2]+ \\
G_{S}\big(u_{j,\phi}^{2}(R,\frac{\pi}{2},\frac{\pi}{6})\big)[1-\cos qD\sqrt{3}/2].
\end{cases}
\end{aligned}
\end{equation}

\noindent Equating the kinetic and potential energies, we get

\begin{equation}
\label{eq16}
\footnotesize
\begin{aligned}
& \omega^{2}=\omega_{1}^{2}+\frac{4}{M_{0}} \\
& \begin{cases}
G_{N}S_{N}\big(1+\cos^{2}{m\frac{\pi}{6}} \big[1+\cos{qD\sqrt{3}/2}\big]\big)+ & \quad m=\text{even}\\
G_{S}S_{S,1}\sin^{2}{m\frac{\pi}{6}} \big[1+\cos{qD\sqrt{3}/2}\big]+\\
G_{S}S_{S,2}\cos^{2}{m\frac{\pi}{6}}\big[1-\cos{qD\sqrt{3}/2}\big],   \\
G_{N}S_{N}\cos^{2}{m\frac{\pi}{6}}\big[1-\cos qD\sqrt{3}/2\big]+ & \quad m=\text{odd}\\
G_{S}S_{S,1}\sin^{2}{m\frac{\pi}{6}}\big[1-\cos qD\sqrt{3}/2\big]+\\
G_{S}S_{S,2}\cos^{2}{m\frac{\pi}{6}}\big[1+\cos{qD\sqrt{3}/2}\big] ,
\end{cases}
\end{aligned}
\end{equation}

\begin{equation}
\footnotesize
\label{eq17}
\begin{aligned}
& S_{N}= M_{0}\frac{u_{r,L,m}^{2}(R,\frac{\pi}{2},0)}{M_{L,m}A^{2}}, \\
& S_{S,1}= M_{0}\frac{u_{\phi,L,m}^{2}(R,\frac{\pi}{2},\frac{\pi}{2})}{M_{L,m}A^{2}}\\
& S_{S,2}= M_{0} \frac{u_{\theta,L,m}^{2}(R,\frac{\pi}{2},0)}{M_{L,m}A^{2}}.
\end{aligned}
\end{equation}

By analyzing Eq. \ref{eq17}, we see that similarly to the case of sphere-substrate contact, 
the properties of the associated Legendre polynomials (Appendix C) give rise to special cases. $S_{N}$ and $S_{S,1}$ are nonzero only if 
$L+m$ is even and $S_{S,2}$ is nonzero only if $L+m$ is odd. In addition, $S_{S,1}$ is zero if $m=0$. These conditions stem from the 
property $P_{L}^{m}(0)=0$ when $L+m$ is an odd number \cite{Arfken2005} and from the recurrence relation in Eq. \ref{eq12}, 
where $\frac{\partial P_{L}^{m}}{\partial \theta}\big \rvert_{\theta=\pi/2}=0$ when $L+m$ is even.

\begin{figure}[h]
\centering
\graphicspath{ {./} }
\includegraphics[width=220pt]{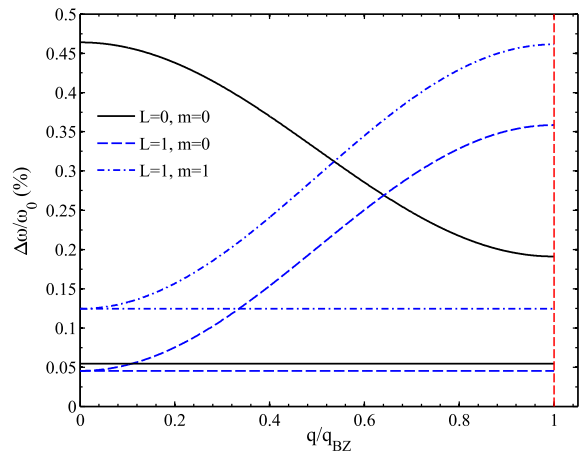}
\caption{Dispersion relation of the spheroidal modes $L=0, 1$, $m=0,\pm1$ calculated using Eqs. \ref{eq16} and \ref{eq17}. 
The horizontal lines only include the effect due to the sphere-substrate contact. The frequency shifts are normalized to the respective 
free-sphere frequencies ($\omega_{0}$).}
\label{fig6}
\end{figure}

Generally, within a mode $L$, the sphere-sphere interaction may mix modes with different values of $m$. This mixing takes place when two or more 
$m$ modes deform the same contact spring. In this case, we will need to find new eigenmodes which will be linear combinations of modes with different $m$ 
values. Such analysis is beyond the scope of the present paper. However, no mode mixing takes place for $L=0$ and L=1. Fig. \ref{fig6} shows 
calculations of the spheroidal dispersion of an hexagonal lattice for the $L=0,1$ modes. We observe that the inclusion of the sphere-sphere interaction 
greatly modifies the spheroidal frequencies compared to the interaction with the substrate alone: it results not only in dispersion, but also in a much larger 
frequency shift.

\section{V. Conclusions }

In summary, we have investigated wave propagation in a hexagonal monolayer of spheres on a substrate and extended the analysis onto 
spheroidal vibrational modes of the spheres. We showed that the hexagonal 
lattice yields three sagittally polarized contact-based modes, similar to the square lattice case; however, there are significant differences in dispersion and the 
motion patterns between the two lattice types. While contact-based vibrational modes originate from Hertzian contacts between the spheres and between 
the spheres and the substrate, spheroidal modes only undergo small perturbations due to Hertzian contacts. We find that the contact with the substrate 
causes a small upshift in the frequency of spheroidal modes with the azimuthal number $m$ equal to zero or $1$. Sphere-sphere contacts cause a 
further frequency upshift as well as dispersion with either positive or negative slope depending on the spheroidal mode. Our model ignored elastic waves 
in the substrate, essentially treating the substrate as rigid (however local elasticity of the substrate at the contact points was accounted for in the Hertzian 
contact model). As has been shown in Ref. \cite{Wallen2015}, the assumption of rigid substrate fails near the intersections of the contact-based modes 
with the Rayleigh wave in the substrate, which results in the hybridization and avoided crossing. It would be straightforward to extend the  analysis of 
this interaction performed in Ref. \cite{Wallen2015} onto the hexagonal lattice case. The interaction of spheroidal modes with the Rayleigh waves in the 
substrate is also expected and presents a subject for the future research. Another expected effect due to acoustic waves in the substrate is the 
attenuation of acoustic modes of the monolayer, whose phase velocity lies above the bulk transverse velocity of the substrate. Quantifying this 
attenuation presents another topic for future investigations. 

\section{Acknowledgments}

The work performed at MIT was supported by the U.S. Department of Energy Grant No. DE-FG02-00ER15087. The work performed 
at the University of Washington was supported by the U.S. National Science Foundation (Grant 
No. CMMI-1333858). The 
work performed at University of Bayreuth was supported by the German Research 
Foundation (DFG, SFB840). The work performed at CINVESTAV was partially supported by projects 192 (``Fronteras de la ciencia'') and 251882 
(``Investigacion Cientifica Basica 2015'') as well as the fund Conacyt-SENER-Energy-Sustainability (Grant No. 207450), within the Strategic Project 
CEMIESol-Cosolpi No. 10 (``Solar Fuels Industrial Proccesses''). A.V-F appreciates support from CINVESTAV 
and CONACYT through normal, mixed scholarships. C.S. acknowledges 
support from the Elite Network Bavaria (ENB).

\section{Appendix A: Contact frequencies for a hexagonal and square lattice}
In the limiting case $q\rightarrow q_{BZ}$, the frequencies for a hexagonal lattice are
\begin{equation}
\label{eq18}
\begin{aligned}
 w^{hex}_{V}&=\sqrt{\bigg(\frac{K_{N}}{\textrm{m}}\bigg)\big(8\xi +1\big)},\\ 
  w^{hex}_{HR}&=\bigg \{ \bigg(\frac{K_{S}}{4\textrm{m}}\bigg)\big(8\sqrt{3}\eta +8 \gamma +7 -\\
& \sqrt{192\eta^{2}+64\gamma^{2}+16\sqrt{3}\eta(8\gamma-3)-48\gamma+49} \big) \bigg \}^{\frac{1}{2}},\\ 
  w^{hex}_{RH}&=\bigg \{ \bigg(\frac{K_{S}}{4\textrm{m}}\bigg)\big(8\sqrt{3}\eta +8 \gamma +7 + \\
&  \sqrt{192\eta^{2}+64\gamma^{2}+16\sqrt{3}\eta(8\gamma-3)-48\gamma+49} \big) \bigg\}^{\frac{1}{2}},
\end{aligned}
\end{equation}

\noindent and for the square lattice the frequencies are 

\begin{equation}
\label{eq19}
\begin{aligned}
 w^{cub}_{V}&=\sqrt{\bigg(\frac{K_{N}}{\textrm{m}}\bigg)\big(4\xi +1\big)},\\ 
  w^{cub}_{HR}&=\bigg \{ \bigg(\frac{K_{S}}{4\textrm{m}}\bigg)\big(8\eta  +7 -\sqrt{64\eta^{2}-48\eta+49} \big) \bigg \}^{\frac{1}{2}},\\ 
  w^{cub}_{HR}&=\bigg \{ \bigg(\frac{K_{S}}{4\textrm{m}}\bigg)\big(8\eta  +7 +\sqrt{64\eta^{2}-48\eta+49} \big) \bigg \}^{\frac{1}{2}}
\end{aligned}
\end{equation}

\noindent where $\xi=G_{S}/K_{N}$, $\eta=G_{N}/K_{S}$, $\gamma=G_{S}/K_{S}$ and m is the microsphere mass.

\section{Appendix B: Spheroidal displacements}

Spheroidal displacement for a stress free sphere (only considering modes symmetric with respect to 
the sagittal plane $x$-$z$) are given by \cite{Eringer1975},

\begin{equation}
\footnotesize
\label{eq20}
\begin{aligned}
& u_{r,L,m}=\frac{A}{r}\bigg( G_{1}(\alpha r)-\frac{T_{11}(\alpha R)}{T_{13}(\beta R)}G_{2}(\beta r) \bigg)P_{L}^{m}(\cos(\theta))\cos{(m\phi)}e^{-i\omega t}, \\
& u_{\theta,L,m}=\frac{A}{r}\bigg( g_{1}(\alpha r) -\frac{T_{11}(\alpha R)}{T_{13}(\beta R)}g_{2}(\beta r) \bigg)\frac{\partial}{\partial \theta}P_{L}^{m}(\cos(\theta))\cos{(m\phi)}e^{-i\omega t}, \\
& u_{\phi,L,m}=-\frac{A}{r}\bigg( g_{1}(\alpha r) -\frac{T_{11}(\alpha R)}{T_{13}(\beta R)} g_{2}(\beta r) \bigg) P_{L}^{m}(\cos(\theta))\bigg(\frac{m\sin (m\phi)}{\sin \theta}\bigg)e^{-i\omega t}, \\
\end{aligned}
\end{equation}

\begin{equation}
\label{eq21}
\begin{aligned}
& G_{1}(x)=Lj_{L}(x)- x j_{L+1}(x), \\
& G_{2}(x) = L(L+1)j_{L}(x), \\
& g_{1}(x)=j_{L}(x), \\
& g_{2}(x)=(L+1)j_{L}(x)-xj_{L+1}(x),
\end{aligned}
\end{equation}

\begin{equation}
\label{eq22}
\begin{aligned}
& T_{11}(x)=\bigg(L^{2}-L-\frac{\beta^{2}R^{2}}{2}\bigg)j_{L}(x)+2\alpha R j_{L+1}(x) ,\\
& T_{13}(x)=L\big(L+1\big)\big[\big(L-1\big)j_{L}(x)-x j_{L+1}(x)\big], \\
& T_{41}(x)=\big(L-1\big)j_{L}(x)-xj_{L+1}(x), \\
& T_{43}(x)=\bigg(L^{2}-1-\frac{1}{2}x^{2}\bigg)j_{L}(x)+xj_{L+1}(x),
\end{aligned}
\end{equation}

\noindent where $A$ is the displacement amplitude, $L$ is the angular number, $m$ is the azimuthal number, 
$\alpha =\omega/c_{L}$, $\beta =\omega/c_{T}$, $P_{L}^{m}$ are 
the associated Legendre polynomials, $j_{L}$ are the spherical Bessel functions, $c_{L}$ and $c_{T}$ are the longitudinal 
and transverse acoustic speeds respectively. 

The characteristic equations for for the vibrations of a free sphere are \cite{Eringer1975},

 \begin{equation}
\label{eq23}
\begin{aligned}
& T_{11}(\alpha R)T_{43}(\beta R)-T_{41}(\alpha R)T_{13}(\beta R)=0, & L>0 \\
& T_{11}(\alpha R)=0, & L=0.
\end{aligned}
\end{equation}

\section{Appendix C: Associated Legendre polynomials}

Rodrigues formula \cite{Arfken2005} for the associated Legendre polynomials is
\begin{equation}
\label{eq24}
\begin{aligned}
P_{L}^{m}(x)=\frac{(-1)^{m}}{2^{L}L!}(1-x^{2})^{\frac{m}{2}}\frac{d^{L+m}}{dx^{L+m}}(x^{2}-1)^{L}.
\end{aligned}
\end{equation}

\noindent In the case of modes with $m<0$, the associated Legendre polynomials can be expressed as 

\begin{equation}
\label{eq25}
\begin{aligned}
P_{L}^{-m}(x)=(-1)^{m}\frac{(L-m)!}{(L+m)!}P_{L}^{m}(x).
\end{aligned}
\end{equation}

We present some important examples relevant to the modes discussed in this work,

\begin{equation}
\label{eq26}
\begin{aligned}
P_{0}^{0}(\cos \theta)&=1,\\
P_{1}^{0}(\cos \theta)&=\cos \theta,\\
P_{1}^{1}(\cos \theta)&=-\sqrt{1-\cos^{2} \theta},\\
P_{2}^{0}(\cos \theta)&=\frac{1}{2}\big(-1+3\cos^{2}\theta \big),\\
P_{2}^{1}(\cos \theta)&=-3\cos \theta \sqrt{1-\cos^{2}\theta},\\
P_{2}^{2}(\cos \theta)&=3\big(1-\cos^{2}\theta \big).
\end{aligned}
\end{equation}

\section{Appendix D: Square lattice spheroidal dispersion}

In the case of a square lattice, the dispersion arising due to sphere-sphere contact can be calculated similarly to the hexagonal lattice. 
Two contact stiffnesses were considered per unit cell, instead of three. The average kinetic and potential energy per oscillation period of the 
unit cell can be expressed as 

\begin{equation}
\label{eq27}
\footnotesize
\begin{aligned}
 <E_{kin}>&=\frac{1}{4}\omega^{2}M_{L,m}A^{2}, & \\
 <E_{pot}>&=\frac{1}{4}K_{L,m}A^{2}  + \\
	& \frac{G_{N}}{4}\big(u_{j,r}(R,\frac{\pi}{2},\frac{\pi}{2})+u_{j,r}(R,\frac{\pi}{2},\frac{3\pi}{2})\big)^{2}+ \\
	 &\frac{G_{N}}{4}\big(u_{j,r}(R,\frac{\pi}{2},0)+u_{j+1,r}(R,\frac{\pi}{2},\pi)\big)^{2}+ \\
	 &\frac{G_{S}}{4}(u_{j,\theta}(R,\frac{\pi}{2},0)-u_{j+1,\theta}(R,\frac{\pi}{2},\pi))^{2},
\end{aligned}
\end{equation}

\noindent further simplification leads to 

\begin{equation}
\label{eq28}
\footnotesize
\begin{aligned}
& <E_{kin}>=\frac{1}{4}\omega^{2}M_{L,m}A^{2}, & \\
& <E_{pot}>=\frac{1}{4}K_{L,m}A^{2}  + \\
& \frac{1}{2}\begin{cases}
	 G_{N}\big(2u_{j,r}^{2}(R,\frac{\pi}{2},\frac{\pi}{2}) +u_{j,r}^{2}(R,\frac{\pi}{2},0)[1+\cos qD]\big) + & \quad m=\text{even}  \\
	 G_{S}u_{j,\theta}^{2}(R,\frac{\pi}{2},0)[1-\cos qD],\\
	 G_{N}u_{j,r}^{2}(R,\frac{\pi}{2},0)[1-\cos qD] +& \quad m=\text{odd}\\
	 G_{S}u_{j,\theta}^{2}(R,\frac{\pi}{2},0)[1+\cos qD].
	\end{cases}
\end{aligned}
\end{equation}

The resulting dispersion is 

\begin{equation}
\label{eq29}
\footnotesize
\begin{aligned}
& \omega^{2}=\omega_{1}^{2}+\frac{2}{M_{0}} \\
& \begin{cases}
G_{N}S_{N}[3+\cos qD] +& \quad m=\text{even}\\
G_{S}S_{S,2}[1-\cos qD], \\
G_{N}S_{N}[1-\cos qD] + & \quad m=\text{odd} \\
G_{S}S_{S,2}(1+\cos qD), 
\end{cases}
\end{aligned}
\end{equation}
 
\noindent where $S_{N}$ and $S_{S,2}$ are given by Eq. \ref{eq17}.

\section{VI. References}


\begin{thebibliography}{21}%
\makeatletter
\providecommand \@ifxundefined [1]{%
 \@ifx{#1\undefined}
}%
\providecommand \@ifnum [1]{%
 \ifnum #1\expandafter \@firstoftwo
 \else \expandafter \@secondoftwo
 \fi
}%
\providecommand \@ifx [1]{%
 \ifx #1\expandafter \@firstoftwo
 \else \expandafter \@secondoftwo
 \fi
}%
\providecommand \natexlab [1]{#1}%
\providecommand \enquote  [1]{``#1''}%
\providecommand \bibnamefont  [1]{#1}%
\providecommand \bibfnamefont [1]{#1}%
\providecommand \citenamefont [1]{#1}%
\providecommand \href@noop [0]{\@secondoftwo}%
\providecommand \href [0]{\begingroup \@sanitize@url \@href}%
\providecommand \@href[1]{\@@startlink{#1}\@@href}%
\providecommand \@@href[1]{\endgroup#1\@@endlink}%
\providecommand \@sanitize@url [0]{\catcode `\\12\catcode `\$12\catcode
  `\&12\catcode `\#12\catcode `\^12\catcode `\_12\catcode `\%12\relax}%
\providecommand \@@startlink[1]{}%
\providecommand \@@endlink[0]{}%
\providecommand \url  [0]{\begingroup\@sanitize@url \@url }%
\providecommand \@url [1]{\endgroup\@href {#1}{\urlprefix }}%
\providecommand \urlprefix  [0]{URL }%
\providecommand \Eprint [0]{\href }%
\providecommand \doibase [0]{http://dx.doi.org/}%
\providecommand \selectlanguage [0]{\@gobble}%
\providecommand \bibinfo  [0]{\@secondoftwo}%
\providecommand \bibfield  [0]{\@secondoftwo}%
\providecommand \translation [1]{[#1]}%
\providecommand \BibitemOpen [0]{}%
\providecommand \bibitemStop [0]{}%
\providecommand \bibitemNoStop [0]{.\EOS\space}%
\providecommand \EOS [0]{\spacefactor3000\relax}%
\providecommand \BibitemShut  [1]{\csname bibitem#1\endcsname}%
\let\auto@bib@innerbib\@empty
%</preamble>
\bibitem [{\citenamefont {Nesterenko}(2001)}]{Nesterenko2001}%
  \BibitemOpen
  \bibfield  {author} {\bibinfo {author} {\bibfnamefont {V.}~\bibnamefont
  {Nesterenko}},\ }\href@noop {} {\emph {\bibinfo {title} {Dynamics of
  heterogeneous materials}}}\ (\bibinfo  {publisher} {Springer Science and
  Business Media.},\ \bibinfo {year} {2001})\BibitemShut {NoStop}%
\bibitem [{\citenamefont {Theocharis}\ \emph {et~al.}(2013)\citenamefont
  {Theocharis}, \citenamefont {Boechler},\ and\ \citenamefont
  {Daraio}}]{Theocharis2013}%
  \BibitemOpen
  \bibfield  {author} {\bibinfo {author} {\bibfnamefont {G.}~\bibnamefont
  {Theocharis}}, \bibinfo {author} {\bibfnamefont {N.}~\bibnamefont
  {Boechler}}, \ and\ \bibinfo {author} {\bibfnamefont {C.}~\bibnamefont
  {Daraio}},\ }\href@noop {} {\emph {\bibinfo {title} {Nonlinear periodic
  phononic structures and granular crystals. In Acoustic Metamaterials and
  Phononic Crystals.}}}\ (\bibinfo  {publisher} {Springer Berlin Heidelberg},\
  \bibinfo {year} {2013})\BibitemShut {NoStop}%
\bibitem [{\citenamefont {Vogel}\ \emph {et~al.}(2015)\citenamefont {Vogel},
  \citenamefont {Retsch}, \citenamefont {Fustin}, \citenamefont {del Campo},\
  and\ \citenamefont {Jonas}}]{Vogel2015}%
  \BibitemOpen
  \bibfield  {author} {\bibinfo {author} {\bibfnamefont {N.}~\bibnamefont
  {Vogel}}, \bibinfo {author} {\bibfnamefont {M.}~\bibnamefont {Retsch}},
  \bibinfo {author} {\bibfnamefont {C.-A.}\ \bibnamefont {Fustin}}, \bibinfo
  {author} {\bibfnamefont {A.}~\bibnamefont {del Campo}}, \ and\ \bibinfo
  {author} {\bibfnamefont {U.}~\bibnamefont {Jonas}},\ }\href@noop {}
  {\bibfield  {journal} {\bibinfo  {journal} {Chem. Rev.}\ }\textbf {\bibinfo
  {volume} {115}},\ \bibinfo {pages} {6265} (\bibinfo {year}
  {2015})}\BibitemShut {NoStop}%
\bibitem [{\citenamefont {Boechler}\ \emph {et~al.}(2013)\citenamefont
  {Boechler}, \citenamefont {Eliason}, \citenamefont {Kumar}, \citenamefont
  {Maznev}, \citenamefont {Nelson},\ and\ \citenamefont {Fang}}]{Boechler2013}%
  \BibitemOpen
  \bibfield  {author} {\bibinfo {author} {\bibfnamefont {N.}~\bibnamefont
  {Boechler}}, \bibinfo {author} {\bibfnamefont {J.~K.}\ \bibnamefont
  {Eliason}}, \bibinfo {author} {\bibfnamefont {A.}~\bibnamefont {Kumar}},
  \bibinfo {author} {\bibfnamefont {A.~A.}\ \bibnamefont {Maznev}}, \bibinfo
  {author} {\bibfnamefont {K.~A.}\ \bibnamefont {Nelson}}, \ and\ \bibinfo
  {author} {\bibfnamefont {N.}~\bibnamefont {Fang}},\ }\href {\doibase
  10.1103/PhysRevLett.111.036103} {\bibfield  {journal} {\bibinfo  {journal}
  {Phys. Rev. Lett.}\ }\textbf {\bibinfo {volume} {111}},\ \bibinfo {pages}
  {036103} (\bibinfo {year} {2013})}\BibitemShut {NoStop}%
\bibitem [{\citenamefont {Khanolkar}\ \emph {et~al.}(2015)\citenamefont
  {Khanolkar}, \citenamefont {Wallen}, \citenamefont {{Abi Ghanem}},
  \citenamefont {Jenks}, \citenamefont {Vogel},\ and\ \citenamefont
  {Boechler}}]{Khanolkar2015}%
  \BibitemOpen
  \bibfield  {author} {\bibinfo {author} {\bibfnamefont {A.}~\bibnamefont
  {Khanolkar}}, \bibinfo {author} {\bibfnamefont {S.}~\bibnamefont {Wallen}},
  \bibinfo {author} {\bibfnamefont {M.}~\bibnamefont {{Abi Ghanem}}}, \bibinfo
  {author} {\bibfnamefont {J.}~\bibnamefont {Jenks}}, \bibinfo {author}
  {\bibfnamefont {N.}~\bibnamefont {Vogel}}, \ and\ \bibinfo {author}
  {\bibfnamefont {N.}~\bibnamefont {Boechler}},\ }\href {\doibase
  10.1063/1.4928564} {\bibfield  {journal} {\bibinfo  {journal} {App. Phys.
  Lett.}\ }\textbf {\bibinfo {volume} {107}},\ \bibinfo {pages} {071903}
  (\bibinfo {year} {2015})}\BibitemShut {NoStop}%
\bibitem [{\citenamefont {Hiraiwa}\ \emph {et~al.}(2016)\citenamefont
  {Hiraiwa}, \citenamefont {{Abi Ghanem}}, \citenamefont {Wallen},
  \citenamefont {Khanolkar}, \citenamefont {Maznev},\ and\ \citenamefont
  {Boechler}}]{Hiraiwa2016}%
  \BibitemOpen
  \bibfield  {author} {\bibinfo {author} {\bibfnamefont {M.}~\bibnamefont
  {Hiraiwa}}, \bibinfo {author} {\bibfnamefont {M.}~\bibnamefont {{Abi
  Ghanem}}}, \bibinfo {author} {\bibfnamefont {S.}~\bibnamefont {Wallen}},
  \bibinfo {author} {\bibfnamefont {A.}~\bibnamefont {Khanolkar}}, \bibinfo
  {author} {\bibfnamefont {A.}~\bibnamefont {Maznev}}, \ and\ \bibinfo {author}
  {\bibfnamefont {N.}~\bibnamefont {Boechler}},\ }\href {\doibase
  10.1103/PhysRevLett.116.198001} {\bibfield  {journal} {\bibinfo  {journal}
  {Phys. Rev. Lett.}\ }\textbf {\bibinfo {volume} {116}},\ \bibinfo {pages}
  {198001} (\bibinfo {year} {2016})}\BibitemShut {NoStop}%
\bibitem [{\citenamefont {Eliason}\ \emph {et~al.}(2016)\citenamefont
  {Eliason}, \citenamefont {Vega-Flick}, \citenamefont {Hiraiwa}, \citenamefont
  {Khanolkar}, \citenamefont {Gan}, \citenamefont {Boechler}, \citenamefont
  {Fang}, \citenamefont {Nelson},\ and\ \citenamefont {Maznev}}]{Eliason2016}%
  \BibitemOpen
  \bibfield  {author} {\bibinfo {author} {\bibfnamefont {J.~K.}\ \bibnamefont
  {Eliason}}, \bibinfo {author} {\bibfnamefont {A.}~\bibnamefont {Vega-Flick}},
  \bibinfo {author} {\bibfnamefont {M.}~\bibnamefont {Hiraiwa}}, \bibinfo
  {author} {\bibfnamefont {A.}~\bibnamefont {Khanolkar}}, \bibinfo {author}
  {\bibfnamefont {T.}~\bibnamefont {Gan}}, \bibinfo {author} {\bibfnamefont
  {N.}~\bibnamefont {Boechler}}, \bibinfo {author} {\bibfnamefont
  {N.}~\bibnamefont {Fang}}, \bibinfo {author} {\bibfnamefont {K.~a.}\
  \bibnamefont {Nelson}}, \ and\ \bibinfo {author} {\bibfnamefont {A.~A.}\
  \bibnamefont {Maznev}},\ }\href {\doibase 10.1063/1.4941808} {\bibfield
  {journal} {\bibinfo  {journal} {App. Phys. Lett.}\ }\textbf {\bibinfo
  {volume} {108}},\ \bibinfo {pages} {061907} (\bibinfo {year}
  {2016})}\BibitemShut {NoStop}%
\bibitem [{\citenamefont {Tournat}\ \emph {et~al.}(2011)\citenamefont
  {Tournat}, \citenamefont {P{\'e}rez-Arjona}, \citenamefont {Merkel},
  \citenamefont {Sanchez-Morcillo},\ and\ \citenamefont {Gusev}}]{Tournat2011}%
  \BibitemOpen
  \bibfield  {author} {\bibinfo {author} {\bibfnamefont {V.}~\bibnamefont
  {Tournat}}, \bibinfo {author} {\bibfnamefont {I.}~\bibnamefont
  {P{\'e}rez-Arjona}}, \bibinfo {author} {\bibfnamefont {A.}~\bibnamefont
  {Merkel}}, \bibinfo {author} {\bibfnamefont {V.}~\bibnamefont
  {Sanchez-Morcillo}}, \ and\ \bibinfo {author} {\bibfnamefont
  {V.}~\bibnamefont {Gusev}},\ }\href@noop {} {\bibfield  {journal} {\bibinfo
  {journal} {New J. Phys.}\ }\textbf {\bibinfo {volume} {13}},\ \bibinfo
  {pages} {073042} (\bibinfo {year} {2011})}\BibitemShut {NoStop}%
\bibitem [{\citenamefont {Wallen}\ \emph {et~al.}(2015)\citenamefont {Wallen},
  \citenamefont {Maznev},\ and\ \citenamefont {Boechler}}]{Wallen2015}%
  \BibitemOpen
  \bibfield  {author} {\bibinfo {author} {\bibfnamefont {S.~P.}\ \bibnamefont
  {Wallen}}, \bibinfo {author} {\bibfnamefont {A.~A.}\ \bibnamefont {Maznev}},
  \ and\ \bibinfo {author} {\bibfnamefont {N.}~\bibnamefont {Boechler}},\
  }\href {\doibase 10.1103/PhysRevB.92.174303} {\bibfield  {journal} {\bibinfo
  {journal} {Phys. Rev. B}\ }\textbf {\bibinfo {volume} {92}},\ \bibinfo
  {pages} {174303} (\bibinfo {year} {2015})}\BibitemShut {NoStop}%
\bibitem [{\citenamefont {Cheng}\ \emph {et~al.}(2006)\citenamefont {Cheng},
  \citenamefont {Wang}, \citenamefont {Jonas}, \citenamefont {Fytas},\ and\
  \citenamefont {Stefanou}}]{Cheng2006}%
  \BibitemOpen
  \bibfield  {author} {\bibinfo {author} {\bibfnamefont {W.}~\bibnamefont
  {Cheng}}, \bibinfo {author} {\bibfnamefont {J.}~\bibnamefont {Wang}},
  \bibinfo {author} {\bibfnamefont {U.}~\bibnamefont {Jonas}}, \bibinfo
  {author} {\bibfnamefont {G.}~\bibnamefont {Fytas}}, \ and\ \bibinfo {author}
  {\bibfnamefont {N.}~\bibnamefont {Stefanou}},\ }\href {\doibase
  10.1038/nmat1727} {\bibfield  {journal} {\bibinfo  {journal} {Nat. Mater.}\
  }\textbf {\bibinfo {volume} {5}},\ \bibinfo {pages} {830} (\bibinfo {year}
  {2006})}\BibitemShut {NoStop}%
\bibitem [{\citenamefont {Tian}(2004)}]{Tian2004}%
  \BibitemOpen
  \bibfield  {author} {\bibinfo {author} {\bibfnamefont {J.}~\bibnamefont
  {Tian}},\ }\href {\doibase 10.1063/1.1737472} {\bibfield  {journal} {\bibinfo
   {journal} {J. App. Phys.}\ }\textbf {\bibinfo {volume} {95}},\ \bibinfo
  {pages} {8366} (\bibinfo {year} {2004})}\BibitemShut {NoStop}%
\bibitem [{\citenamefont {Hladky-Hennion}\ \emph {et~al.}(2002)\citenamefont
  {Hladky-Hennion}, \citenamefont {Cohen-Tenoudji}, \citenamefont {Devos},\
  and\ \citenamefont {de~Billy}}]{Hladky-Hennion2002}%
  \BibitemOpen
  \bibfield  {author} {\bibinfo {author} {\bibfnamefont {A.-C.}\ \bibnamefont
  {Hladky-Hennion}}, \bibinfo {author} {\bibfnamefont {F.}~\bibnamefont
  {Cohen-Tenoudji}}, \bibinfo {author} {\bibfnamefont {A.}~\bibnamefont
  {Devos}}, \ and\ \bibinfo {author} {\bibfnamefont {M.}~\bibnamefont
  {de~Billy}},\ }\href {\doibase 10.1121/1.1497369} {\bibfield  {journal}
  {\bibinfo  {journal} {J. Acoust. Soc. Am.}\ }\textbf {\bibinfo {volume}
  {112}},\ \bibinfo {pages} {850} (\bibinfo {year} {2002})}\BibitemShut
  {NoStop}%
\bibitem [{\citenamefont {Hladky-Hennion}\ \emph {et~al.}(2004)\citenamefont
  {Hladky-Hennion}, \citenamefont {Devos},\ and\ \citenamefont
  {de~Billy}}]{Hladky-Hennion2004}%
  \BibitemOpen
  \bibfield  {author} {\bibinfo {author} {\bibfnamefont {A.-C.}\ \bibnamefont
  {Hladky-Hennion}}, \bibinfo {author} {\bibfnamefont {A.}~\bibnamefont
  {Devos}}, \ and\ \bibinfo {author} {\bibfnamefont {M.}~\bibnamefont
  {de~Billy}},\ }\href {\doibase 10.1121/1.1763598} {\bibfield  {journal}
  {\bibinfo  {journal} {J. Acoust. Soc. Am.}\ }\textbf {\bibinfo {volume}
  {116}},\ \bibinfo {pages} {117} (\bibinfo {year} {2004})}\BibitemShut
  {NoStop}%
\bibitem [{\citenamefont {Johnson}(1987)}]{Johnson1987}%
  \BibitemOpen
  \bibfield  {author} {\bibinfo {author} {\bibfnamefont {K.~L.}\ \bibnamefont
  {Johnson}},\ }\href@noop {} {\emph {\bibinfo {title} {Contact mechanics}}}\
  (\bibinfo  {publisher} {Cambridge university press},\ \bibinfo {year}
  {1987})\BibitemShut {NoStop}%
\bibitem [{\citenamefont {Muller}\ \emph {et~al.}(1983)\citenamefont {Muller},
  \citenamefont {Dejarguin},\ and\ \citenamefont {Toporov}}]{Muller1983}%
  \BibitemOpen
  \bibfield  {author} {\bibinfo {author} {\bibfnamefont {V.~M.}\ \bibnamefont
  {Muller}}, \bibinfo {author} {\bibfnamefont {B.~V.}\ \bibnamefont
  {Dejarguin}}, \ and\ \bibinfo {author} {\bibfnamefont {Y.~P.}\ \bibnamefont
  {Toporov}},\ }\href@noop {} {\bibfield  {journal} {\bibinfo  {journal}
  {Colloids and Surf.}\ }\textbf {\bibinfo {volume} {7}},\ \bibinfo {pages}
  {251} (\bibinfo {year} {1983})}\BibitemShut {NoStop}%
\bibitem [{\citenamefont {Israelachvili}(2011)}]{Israelachvili2011}%
  \BibitemOpen
  \bibfield  {author} {\bibinfo {author} {\bibfnamefont {J.}~\bibnamefont
  {Israelachvili}},\ }\href@noop {} {\emph {\bibinfo {title} {Intermoleculat
  and Surface Forces}}},\ edited by\ \bibinfo {editor} {\bibfnamefont
  {I.}~\bibnamefont {Elsevier}}\ (\bibinfo {address} {Burlington, MA},\
  \bibinfo {year} {2011})\BibitemShut {NoStop}%
\bibitem [{\citenamefont {Lamb}(1881)}]{Lamb1881}%
  \BibitemOpen
  \bibfield  {author} {\bibinfo {author} {\bibfnamefont {H.}~\bibnamefont
  {Lamb}},\ }\href@noop {} {\bibfield  {journal} {\bibinfo  {journal} {P. Lond.
  Math. Soc.}\ }\textbf {\bibinfo {volume} {1}},\ \bibinfo {pages} {189}
  (\bibinfo {year} {1881})}\BibitemShut {NoStop}%
\bibitem [{\citenamefont {Eringen}\ and\ \citenamefont
  {Suhubi}(1975)}]{Eringer1975}%
  \BibitemOpen
  \bibfield  {author} {\bibinfo {author} {\bibfnamefont {A.~C.}\ \bibnamefont
  {Eringen}}\ and\ \bibinfo {author} {\bibfnamefont {E.~S.}\ \bibnamefont
  {Suhubi}},\ }\href@noop {} {\emph {\bibinfo {title} {Elastodynamics, Volume
  II Linear Theory}}}\ (\bibinfo {address} {Academic Press, New York},\
  \bibinfo {year} {1975})\BibitemShut {NoStop}%
\bibitem [{\citenamefont {Dehoux}\ \emph {et~al.}(2009)\citenamefont {Dehoux},
  \citenamefont {Kelf}, \citenamefont {Tomoda}, \citenamefont {Matsuda},
  \citenamefont {Wright}, \citenamefont {Ueno}, \citenamefont {Nishijima},
  \citenamefont {Juodkazis}, \citenamefont {Misawa}, \citenamefont {Tournat},\
  and\ \citenamefont {Gusev}}]{Dehoux2009}%
  \BibitemOpen
  \bibfield  {author} {\bibinfo {author} {\bibfnamefont {T.}~\bibnamefont
  {Dehoux}}, \bibinfo {author} {\bibfnamefont {T.~A.}\ \bibnamefont {Kelf}},
  \bibinfo {author} {\bibfnamefont {M.}~\bibnamefont {Tomoda}}, \bibinfo
  {author} {\bibfnamefont {O.}~\bibnamefont {Matsuda}}, \bibinfo {author}
  {\bibfnamefont {O.~B.}\ \bibnamefont {Wright}}, \bibinfo {author}
  {\bibfnamefont {K.}~\bibnamefont {Ueno}}, \bibinfo {author} {\bibfnamefont
  {Y.}~\bibnamefont {Nishijima}}, \bibinfo {author} {\bibfnamefont
  {S.}~\bibnamefont {Juodkazis}}, \bibinfo {author} {\bibfnamefont
  {H.}~\bibnamefont {Misawa}}, \bibinfo {author} {\bibfnamefont
  {V.}~\bibnamefont {Tournat}}, \ and\ \bibinfo {author} {\bibfnamefont
  {V.~E.}\ \bibnamefont {Gusev}},\ }\href
  {http://www.ncbi.nlm.nih.gov/pubmed/19953180} {\bibfield  {journal} {\bibinfo
   {journal} {Opt. Lett.}\ }\textbf {\bibinfo {volume} {34}},\ \bibinfo {pages}
  {3740} (\bibinfo {year} {2009})}\BibitemShut {NoStop}%
\bibitem [{\citenamefont {Mead}(1973)}]{Mead1973}%
  \BibitemOpen
  \bibfield  {author} {\bibinfo {author} {\bibfnamefont {D.~J.}\ \bibnamefont
  {Mead}},\ }\href@noop {} {\bibfield  {journal} {\bibinfo  {journal} {J. Sound
  Vid.}\ }\textbf {\bibinfo {volume} {27}},\ \bibinfo {pages} {235} (\bibinfo
  {year} {1973})}\BibitemShut {NoStop}%
\bibitem [{\citenamefont {Arfken}\ and\ \citenamefont
  {Hans}(2005)}]{Arfken2005}%
  \BibitemOpen
  \bibfield  {author} {\bibinfo {author} {\bibfnamefont {G.~B.}\ \bibnamefont
  {Arfken}}\ and\ \bibinfo {author} {\bibfnamefont {J.~W.}\ \bibnamefont
  {Hans}},\ }\href@noop {} {\emph {\bibinfo {title} {Mathematical methods for
  physicists international student edition}}},\ edited by\ \bibinfo {editor}
  {\bibfnamefont {A.}~\bibnamefont {Press}}\ (\bibinfo {year}
  {2005})\BibitemShut {NoStop}%
\end{thebibliography}
\end{document}